\documentclass[]{aa} 
\usepackage{txfonts}
\usepackage{epsfig}
\usepackage{graphicx}
\usepackage{url}
\usepackage{algorithm2e}
\usepackage{amssymb}
\usepackage{color}
\usepackage[breaklinks=true,colorlinks=true,urlcolor=blue,citecolor=blue,linkcolor=blue]{hyperref}

\newcommand{\epsilonbold}{\mbox{\boldmath$\epsilon$}}

\newcommand{\alphabold}{\mbox{\boldmath$\alpha$}}

\newcommand{\hazel}{\textsc{Hazel}}
\newcommand{\argmin}{\mathop{\mathrm{argmin}}\limits}

\usepackage{natbib,twoopt}
\bibpunct{(}{)}{;}{a}{}{,}             
\makeatletter
  \newcommandtwoopt{\citeads}[3][][]{\href{http://adsabs.harvard.edu/abs/#3}%
    {\def\hyper@linkstart##1##2{}%
     \let\hyper@linkend\@empty\citealp[#1][#2]{#3}}}
  \newcommandtwoopt{\citepads}[3][][]{\href{http://adsabs.harvard.edu/abs/#3}%
    {\def\hyper@linkstart##1##2{}%
     \let\hyper@linkend\@empty\citep[#1][#2]{#3}}}
  \newcommandtwoopt{\citetads}[3][][]{\href{http://adsabs.harvard.edu/abs/#3}%
    {\def\hyper@linkstart##1##2{}%
     \let\hyper@linkend\@empty\citet[#1][#2]{#3}}}
  \newcommandtwoopt{\citeyearads}[3][][]%
    {\href{http://adsabs.harvard.edu/abs/#3}
    {\def\hyper@linkstart##1##2{}%
     \let\hyper@linkend\@empty\citeyear[#1][#2]{#3}}}
\makeatother

\begin{document}

\title{Inversion of Stokes Profiles with Systematic Effects}

\author{A. Asensio Ramos\inst{1,2}, J. de la Cruz Rodr\'{\i}guez\inst{3}, M. J. Mart\'{\i}nez Gonz\'alez\inst{1,2}, \and A. Pastor Yabar\inst{1,2}}

\institute{
 Instituto de Astrof\'{\i}sica de Canarias, 38205, La Laguna, Tenerife, Spain; \email{aasensio@iac.es}
\and
Departamento de Astrof\'{\i}sica, Universidad de La Laguna, E-38205 La Laguna, Tenerife, Spain
\and
Institute for Solar Physics, Dept. of Astronomy, Stockholm University, Albanova University Center, 10691 Stockholm, Sweden
\email{jaime@astro.su.se}
}
             
  \date{Received ---; accepted ---} 

  \abstract{Quantitative thermodynamical, dynamical and magnetic properties of the solar and stellar plasmas are obtained
  by interpreting their emergent non-polarized and polarized spectrum. This inference requires the selection of a set of 
  spectral lines particularly sensitive to the physical conditions in the plasma and a suitable parametric model of the
  solar/stellar atmosphere. Nonlinear inversion codes are then used to fit the model to the observations. However, the
  presence of systematic effects like nearby or blended spectral lines, telluric absorption or incorrect correction
  of the continuum, among others, can strongly affect the results. We present an extension to current inversion codes that
  can deal with these effects in a transparent way. The resulting algorithm is very simple and can be applied to any
  existing inversion code with the addition of a few lines of code as an extra step in each iteration.}

   \authorrunning{Asensio Ramos et al.}
   \titlerunning{Inversion of Stokes Profiles with Systematic Effects}
   \maketitle
%

\section{Introduction}
The decade of 1970's will be remembered as the one in which Solar Physicist were able to really start to infer the
magnetic and thermodynamic properties of the solar plasma from the observations. At that specific time, there was a 
sweet coincidence. On the one hand, the theory of radiative transfer for polarized light was already in its maturity. On
the other, computers started to be available for researchers in general and powerful enough to carry out complex calculations.
It was then the time at which the ideas of non-linear inversion codes were set \citep{harvey72,auer_heasly_house77,skumanich_lites87}.

Inversion algorithms are able to extract information about the magnetic and thermodynamical properties of the solar plasma from the analysis of spectropolarimetric observations.
They function by proposing a specific model to explain the observations and then defining a merit function (usually the $\chi^2$ function, 
valid under the presence of uncorrelated Gaussian noise). The model parameters are iteratively modified for optimizing the merit
function. The first inversion codes were relatively simple and based on strongly simplifying assumptions, like the 
Milne-Eddington (ME) approximation to analytically solve the radiative transfer equation \citep[e.g.,][]{harvey72,auer_heasly_house77,landi_landolfi04}.
Such inversion codes are still used today, like VFISV \citep{borrero07,borrero_vfisv10}, MILOS \citep{orozco_hinode07} or
MERLIN \citep{skumanich_lites87,lites07}.

An enormous step forward was introduced by \cite{sir92}, who developed SIR (Stokes Inversion based on Response functions), an inversion
code that recovers the optical depth stratification of the physical quantities (temperature, magnetic field, velocity, etc.)
from the interpretation of the Stokes profiles. These codes are based on the idea of response function \citep{landi_response77}, that
allows the user to link the perturbation in the emergent Stokes parameters with perturbations in the physical parameters. One of the
key ingredients that facilitated the development of such inversion codes was the possibility to find an analytical
expression for the response functions in local thermodynamic equilibrium (LTE) \citep{sanchez_almeida92,ruizcobo_deltoro94}.
Based on the seed of SIR, several non-linear inversion codes are now available, some of them even dealing with the
much more difficult case of the inversion of spectral lines in non-LTE \citep[NICOLE;][]{socas_trujillo_ruiz00,socas_nicole14}.
All these 1D inversion codes are based on the concept of ``nodes'', that need to be defined a-priori. These nodes mark positions
along the optical depth axis where the value of the physical parameters will be modified to fit the Stokes profiles. The 
full stratification of the atmosphere between the nodes, which is needed
to accurately integrate the radiative transfer equation and to derive the gas pressure scale, is interpolated using a piece-wise polynomial. 
The complexity of the solution then critically
depends on the number of nodes that are employed to describe each of the physical parameters.

After more than a decade without any fundamental improvement \citep[except perhaps the introduction of Bayesian inference
into the field;][]{asensio_martinez_rubino07,asensio_hinode09,asensioramos_modcomp12}, we are nowadays living another
sweet era, again driven by the improvements in computational power. On one side, \cite{vannoort12} has developed a spatially
coupled two-dimensional inversion code in which the effect of the telescope point spread function (PSF) is taken into account.
The PSF couples nearby pixels so that deconvolution and inversion is done at the same time. Following a similar motivation,
\cite{ruizcobo_asensioramos13} have used a regularized deconvolution of the Stokes profiles based on the principal
component analysis (PCA) and have used SIR to invert the deconvolved Stokes parameters. 

\begin{figure*}
\includegraphics[width=\textwidth]{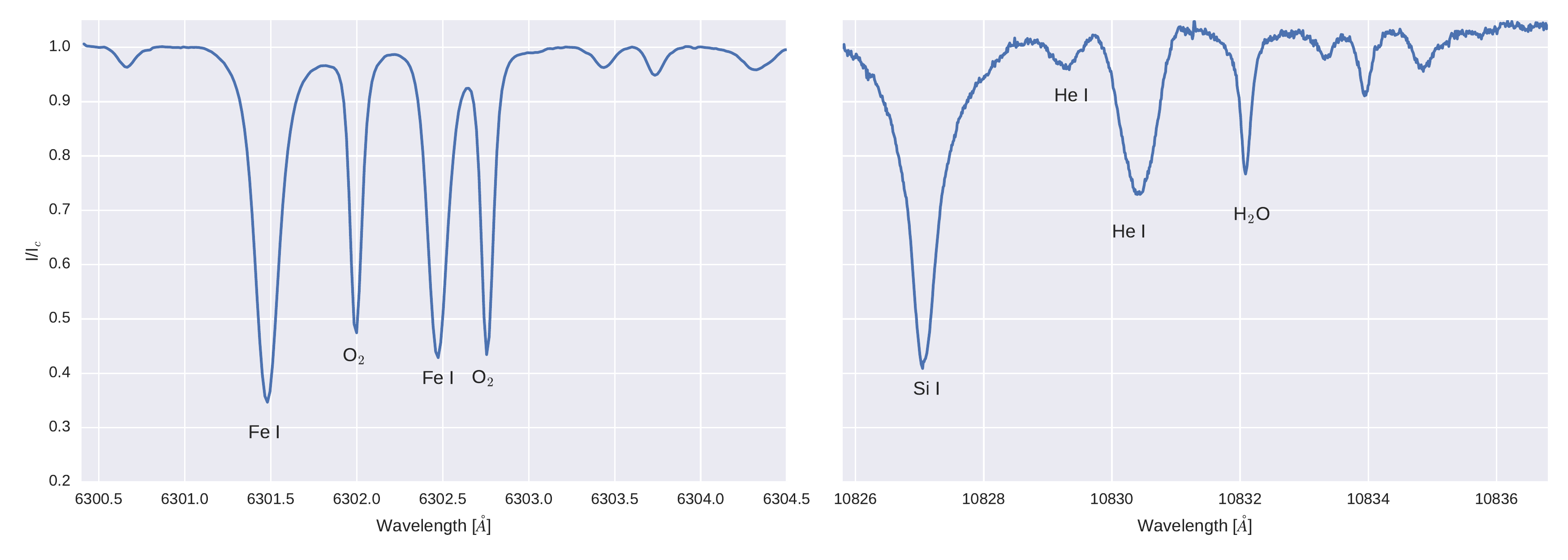}
\caption{Examples of situations in which systematic effects are important when extracting information from
spectral lines.}
\label{fig:examples}
\end{figure*}

Arguably, the last step in the evolution of inversion codes has been the introduction of regularization ideas based on the
concept of sparsity or compressibility\footnote{Given an $n$-dimensional vector $\mathbf{x}$, sparsity means that the majority of the elements of the vector are
strictly zero, while compressibility means that the elements of such vector, when ordered by absolute value, fulfill $|x_i| \leq C i^{-1/r}$.} 
\cite[e.g.,][]{starck10}. \cite{asensio_delacruz15} have presented an inversion code
for the inversion of Stokes profiles that uses $\ell_0$ or $\ell_1$ regularization on a transformed space\footnote{The $\ell_p$-norm
is given by: $\Vert \mathbf{x} \Vert_p = \left( \sum_i |x_i|^p \right) ^{1/p}$, with $p\geq0$.}. The two-dimensional
maps of parameters are linearly transformed (Fourier, wavelet or any other appropriate transformation can be used) and assumed to
be sparse in the transformed domain. This introduces two important constraints. First, the sparsity assumption reduces the number
of unknowns in the problem, avoiding overfitting. Second, the global character of the transformations that are routinely used, spatial
correlation of the result is automatically taken into account. Consequently, the Stokes parameters observed at every pixel potentially 
introduces constraints onto every other pixel of the observed map. This is the first time that the inherent spatial correlation of
the physical parameters has been taken into account in inversion codes. This results into much more stable inversion codes that 
do not produce spurious pixel-by-pixel variations of the maps of physical quantities. Additionally, the complexity of the solution
is automatically adapted, producing more structure where it is needed.

In the conclusions of \cite{asensio_delacruz15} we pointed out that the sparsity regularization can be applied to invert
Stokes profiles with systematic effects. This is precisely what we present in this paper. 
A customary way of dealing with systematic effects in current inversions is to downweight the influence
on the merit function of the parts of the spectrum that are affected by these effects. Although it works in
practice, it depends on a set of parameters (e.g., region and factor of the downweighting) that make it
quite subjective. Instead, we propose several basis sets
(orthogonal and non-orthogonal) to absorb the systematic effects (for instance, telluric lines) in the Stokes profiles and 
use a proximal projection algorithm \citep[e.g.,][]{parikh_boyd14} to make the solution automatically adapt to the necessary complexity.
We will show that the modifications needed in existing inversion codes are minimal so that
this approach can be introduced without much effort.

\section{Sparsity regularization}
Our objective is to fit the observed Stokes parameters, that are discretized at $N$ finite wavelength 
points $\lambda_j$. For simplicity, we stack the four Stokes parameters in a long vector 
of length $4N$ so that
$\mathbf{O}=\left[I(\lambda_1),\ldots,I(\lambda_N),Q(\lambda_1),\ldots,U(\lambda_1),\ldots,V(\lambda_1),\ldots,V(\lambda_N)\right]^\dag$ 
(with $\dag$ the transpose). The fit is carried out using a model atmosphere, with the aim of extracting useful 
thermal, dynamic and magnetic information
from them. Additionally, we make the assumption that these Stokes parameters are perturbed by some uncontrollable
systematic effects. These systematic effects can be, for instance, telluric lines produced by absorption
in the Earth atmosphere, variations along the spectral direction produced by an incorrect illumination of the camera or
a low-quality flatfielding, etc. Examples of these situations are found in Fig. \ref{fig:examples}. The left panel shows the
well-known region around 6302 \AA\, that contains two \ion{Fe}{i} lines, which are blended with two telluric lines. This
is of special importance for the Fe \textsc{i} line at 6302.5 \AA. Extracting physical information from the line
requires then to avoid the far wings, which can lead to problems in the deepest parts of the atmosphere. The 
right panel shows the region around 10830 \AA, containing a \ion{He}{i} multiplet used for chromospheric diagnostics, which
is blended with a photospheric \ion{Si}{i} line and a telluric line. The interpretation of the He \textsc{i} multiplet, given
that it forms on the extended wings of the Si \textsc{i} line, usually requires a previous analysis of the photospheric line.
Additionally, the strong chromospheric dynamics induces that the He \textsc{i} multiplet sometimes blends with 
the telluric absorption.

In very general terms, we can explain the observations with the following generative model:
\begin{equation}
\mathbf{O} = \mathbf{S}_\mathrm{atm}(\mathbf{q}) + \mathbf{D} \alphabold + \epsilonbold,
\label{eq:generative}
\end{equation}
where $\alphabold$ is a vector of dimension $d$ that describes the systematic effects via a dictionary
$\mathbf{D} \in \mathbb{R}^{4N \times d}$, also known as synthesis operator. We remind that a dictionary is just a collection
of (possibly non-orthogonal) functions that are used to describe the systematic effects. The contribution
$\mathbf{S}_\mathrm{atm}(\mathbf{q})$ contains the Stokes parameters that emerge from a model
atmosphere depending on a vector of model parameters $\mathbf{q}$, ordered like in $\mathbf{O}$.
The noise components $\epsilonbold$ are considered to be Gaussian random variables with zero mean and 
diagonal covariance matrix.
Note that when the size of the dictionary equals the number of observed spectral points 
($d=4N$) and $\mathbf{D}= \mathbf{1}$ (with $\mathbf{1}$ the identity matrix), the model for the systematic effects described in 
Eq. (\ref{eq:generative}) turns out to be very flexible because $\alphabold$ refers to the particular value of the systematic effects at each
sampled wavelength point.

From this generative model, the merit function one has to optimize is the $\chi^2$ with diagonal covariance, in our case given by:
\begin{equation}
\chi^2_\mathbf{D}(\mathbf{q},\alphabold) = \frac{1}{4N} \sum_{j=1}^{4N}  w_j
\frac{\left[S_{\mathrm{atm},j}(\mathbf{q})+ (\mathbf{D} \alphabold)_{j}-O_j\right]^2}{\sigma_{j}^2},
\label{eq:chi2_q}
\end{equation}
where we make explicit the dependence of the merit function on the election of the dictionary.
The previous merit function considers a potentially different noise variance $\sigma_{j}^2$ for every
Stokes parameter and wavelength position, which is surely the case for very strong lines in which the number of photons
in the core is much more absorbed than the wings. Additionally, it is customary to introduce weights $w_j$ for each Stokes
parameter to increase the sensitivity to some parameters when carrying out the inversion\footnote{Note that the role
of the weights is to modify the value of the noise variance. If the noise variance is artificially increased for one Stokes
parameter, its constraining power decreases.}. These weights are precisely the ones that are currently used to 
deal with systematic effects, by decreasing their values for specific wavelength points.

The inferred physical parameters are typically found by solving the following maximum-likelihood problem:
\begin{equation}
\argmin_\mathbf{q,\alphabold} \, \chi^2_\mathbf{D}(\mathbf{q},\alphabold),
\label{eq:problem_l2}
\end{equation}
where the operator ``$\argmin$'' returns the value of $\mathbf{q}$ and $\alphabold$ that minimizes the 
$\chi^2_\mathbf{D}(\mathbf{q},\alphabold)$ function. Note that Eq. (\ref{eq:problem_l2}) is, in general, non-convex\footnote{When the
function to minimize is convex in the variables, a local minimum is also the global minimum. This cannot be
guaranteed in the optimization of non-convex functions, which is typically the case in the inversion
of Stokes parameters} for the inversion of Stokes parameters.
Therefore, we only aspire to reach one of the local minima and later check for the physical relevance of the
solution. Problem (\ref{eq:problem_l2}) is usually solved by direct application of the Levenberg-Marquardt (LM) algorithm 
\citep{levenberg44,marquardt63}, which is 
especially suited to the optimization of such $\ell_2$-norms. Although problem (\ref{eq:problem_l2}) is stated without any constraint,
it is clear from the introduction that it is customary to regularize the solution by imposing some regularity of the solution through
the use of nodes. Consequently, the vector of parameters $\mathbf{q}$ contains the value of the physical parameters (temperature, magnetic field, velocity, \ldots)
at a small number of depths in optical depth in the atmosphere. In between these points, the physical properties are
interpolated from the values at these nodes.

Without any additional constraint for $\alphabold$, it is sure that we will encounter cross-talk between
$\alphabold$ and $\mathbf{q}$. The fundamental reason is that our regressor for the systematic effects is so flexible that
it can potentially fit the observations perfectly, including the specific noise realization. As noted above,
if $d=4N$ and $\mathbf{D}=\mathbf{1}$, a solution
to Eq. (\ref{eq:problem_l2}) is $\alphabold=\mathbf{O}$ and $\mathbf{S}_\mathrm{atm}(\mathbf{q})=0$
for every observed wavelength. This trivial solution is of no interest because it does not extract any
relevant physical information from the observations. In order to overcome this possibility, we follow recent 
ideas \citep{candes06,starck10,asensio_delacruz15} and regularize the problem by imposing a
sparsity constraint on $\alphabold$ or, in general, in any transformation of $\alphabold$.
Two fundamental approaches to include this sparsity penalty have been proposed \citep{starck10}. Each 
one has advantages and disadvantages, as we will show in Sec. \ref{sec:examples}.

\subsection{The analysis penalty approach}
This approach is based on having as many degrees of freedom as observed data points ($d=4N$) and $\mathbf{D}=\mathbf{1}$ and solving the
following problem:
\begin{equation}
\argmin_\mathbf{q,\alphabold} \, \chi^2_\mathbf{1}(\mathbf{q},\alphabold),  \,\,
\mathrm{subject\, to\,} \Vert \mathbf{W} \alphabold \Vert_p \leq s,
\label{eq:analysis_problem0}
\end{equation}
where $\mathbf{W}$ is the matrix associated to any linear transformation of interest, either orthogonal or not, while
$s$ is a predefined threshold. We also remind the reader that $\Vert \mathbf{x} \Vert_p = \left( \sum_i |x_i|^p \right) ^{1/p}$
is the $\ell_q$-norm. For instance, the $\ell_0$ norm of a vector is just the number of non-zero elements, while
the $\ell_1$ norm is the sum of the absolute value of its elements.
Put in words, solving Eq. (\ref{eq:analysis_problem0}) requires to seek the pair $(\mathbf{q},\alphabold)$ 
that better fits our observations, imposing that
the projection of the systematic effects on the transformed domain defined by $\mathbf{W}$ is sparse. The
name \emph{analysis penalty} comes from the fact that $\mathbf{W}$ is the analysis operator, that
carries the vector $\alphabold$ to the transformed sparsity-inducing domain \citep{elad07,starck10}. It
is especially suited to deal with cases in which $\mathbf{W}$ is non-orthogonal and/or overcomplete.

The solution to the previous problem when $p=0$ 
\citep[or equivalently when $p=1$ under some conditions;][]{candes06,donoho06} is known to 
coincide with the exact solution when it exists. Sparsity is, therefore, a very convenient
regularization. The only degree of freedom is to find the appropriate transformation $\mathbf{W}$.
The same problem can be equivalently written with the addition of a regularization parameter ($\lambda$):
\begin{equation}
\argmin_\mathbf{q,\alphabold} \, \left[ \frac{1}{4N} \sum_{j=1}^{4N}  w_j
\frac{\left[S_{\mathrm{atm},j}(\mathbf{q})+ \alpha_{j}-O_j\right]^2}{\sigma_{j}^2} + \lambda \Vert \mathbf{W} \alphabold \Vert_p \right].
\label{eq:problem_l0}
\end{equation}

\subsection{The synthesis penalty approach}
This approach is based on having $\mathbf{D}=\mathbf{W}^T$, and imposing 
a sparsity constraint on the $\alphabold$ vector itself. Consequently, we have to solve
the problem
\begin{equation}
\argmin_\mathbf{q,\alphabold} \, \chi^2_\mathbf{W^T}(\mathbf{q},\alphabold),  \,\,
\mathrm{subject\, to\,} \Vert \alphabold \Vert_p \leq s,
\end{equation}
which, in lagrangian form, becomes:
\begin{equation}
\argmin_\mathbf{q,\alphabold} \, \left[ \frac{1}{4N} \sum_{j=1}^{4N}  w_j 
\frac{\left[S_{\mathrm{atm},j}(\mathbf{q})+ (\mathbf{W}^T \alphabold)_{j}-O_j\right]^2}{\sigma_{j}^2} + \lambda \Vert \alphabold \Vert_p \right],
\label{eq:problem_l0_synthesis}
\end{equation}
The name \emph{synthesis penalty} comes from the fact that $\mathbf{W}^T$ is the synthesis operator, that
generates the systematic effects from the vector $\alphabold$ living on the sparsity-inducing transformed domain.

It was noted by \cite{elad07} that both approaches are equivalent when $\mathbf{W}$ is an
orthogonal transform (like Fourier, wavelet, \ldots), because it fulfills $\mathbf{W}^T \mathbf{W}=\mathbf{1}$. However, they solve completely different
problems when $\mathbf{W}$ is non-orthogonal.

\subsection{Transformations}
We will consider in this paper three options for the regularization term.
The first one uses as regularization an orthogonal wavelet transform. The fact that
$\mathbf{W}$ is orthogonal will slightly simplify the algorithms. The second one is a
non-orthogonal overcomplete isotropic undecimated wavelet transform using the B$_3$-spline \citep{starck10}.
Finally, we will use a hand-made non-orthogonal transform made of Voigt functions centered 
at specific locations in the spectrum, that will be used to absorb the systematic effects. We defer the detailed
description of each one until Sec. \ref{sec:proximal}.

\section{The proposed optimization}
Given the special structure of both problems defined in Eqs. (\ref{eq:problem_l0}) and (\ref{eq:problem_l0_synthesis}), 
in which the regularization only occurs for the $\alphabold$ variables,
we propose to use an alternating optimization method. If $\alphabold$ is fixed, Eq. (\ref{eq:problem_l0})
becomes the traditional least-squares problem for $\mathbf{q}$, that is solved efficiently with Newton-type
methods like the Levenberg-Marquardt algorithm. This method uses second-order information given by $\mathbf{H}_q$, the
Hessian of the merit function with respect to the $\mathbf{q}$ variables:
\begin{equation}
\mathbf{q}_{i+1} = \mathbf{q}_i - \hat{\mathbf{H}}_q^{-1} \nabla_q \chi^2_\mathbf{1}(\mathbf{q}_i,\alphabold_i) \qquad \,\,\, 
\qquad \text{with $\alphabold_i$ fixed}.
\label{eq:q_update}
\end{equation}
Conforming to the prescriptions of the Levenberg-Marquardt algorithm, we
use a modified Hessian matrix by enhancing its diagonal by a factor $\beta$, so 
that $\hat{\mathbf{H}}_q=\mathbf{H}_q+ \beta \mathrm{diag}(\mathbf{H}_q)$. 
This hyperparameter is modified during the
iteration to shift between the gradient descent method (large $\beta$) and Newton-type method (small $\beta$).
We note that inverting the Hessian matrix with a truncated singular value decomposition \citep{sir92}
introduces an extra regularization on the physical parameters at the nodes that is often needed.

\subsection{The analysis penalty}\label{sec:prior}
On the other hand, when $\mathbf{q}$ is fixed, Eq. (\ref{eq:problem_l0}) becomes a 
standard sparsity-constrained linear problem for $\alphabold$.
This problem can be solved efficiently using proximal algorithms \citep{parikh_boyd14,asensio_delacruz15}, which 
are especially suited to solve problems of the type
\begin{equation}
\argmin_{\alphabold} f(\alphabold) := g(\alphabold) + h(\alphabold),
\label{eq:problem_general}
\end{equation}
where $g(\alphabold)=\chi^2_\mathbf{1}(\mathbf{q},\alphabold)$, is a smooth function and 
$h(\alphabold)=\lambda \Vert \mathbf{W}\alphabold \Vert_p$ is a convex 
but not necessarily smooth function (note that the derivative of the $\Vert \mathbf{W} \alphabold \Vert_p$ term is not
continuous). We propose the following first-order iterative scheme to solve the problem:
\begin{equation}
\alphabold_{i+1} = \mathrm{prox}_{p,\lambda,\mathbf{W}} \left[\alphabold_i - 
\tau \nabla_{\alpha} \chi^2_\mathbf{1}(\mathbf{q}_{i+1},\alphabold_i) \right] 
\quad \text{with $\mathbf{q}_{i+1}$ fixed},
\label{eq:iteration_alpha1}
\end{equation}
where the operator $\mathrm{prox}_{p,\lambda,\mathbf{W}}$ is the proximal projection operator \citep{parikh_boyd14}
associated with the constraint $\Vert \mathbf{W} \alphabold \Vert_p$, that we show how to efficiently compute in
Sec. \ref{sec:examples}. The election of the step size $\tau$ is important for the convergence
of the algorithm. It is known to converge provided the step size fulfills $\tau < 2 / \Vert \mathbf{H}_\alpha \Vert^2$, where 
$\Vert \mathbf{H}_\alpha \Vert$ is the spectral norm of the Hessian of the merit function with respect to $\alphabold$ (given by the
square root of the maximum eigenvalue of $\mathbf{H}_\alpha^T \mathbf{H}_\alpha$). We note that faster algorithms like 
FISTA \citep{beck_teboulle09,asensio_delacruz15} can also be used.

Given the simple dependence of $\chi^2_{\mathbf{q},\alphabold}$ on $\alphabold$, 
it can be proven \citep[e.g.,][]{starck10} that Eq. (\ref{eq:iteration_alpha1}) can be simplified to finally obtain:
\begin{equation}
\alphabold_{i+1} = \mathrm{prox}_{p,\lambda,\mathbf{W}} \left[ \mathbf{S}_{\mathrm{atm}}(\mathbf{q}_{i+1})-\mathbf{O} \right] 
\quad \text{with $\mathbf{q}_{i+1}$ fixed}.
\label{eq:iteration_alpha}
\end{equation}
In other words, the estimation of the systematic effects for a new iteration is very simple and reduces to
computing the proximal projection of the residual between the observed Stokes profiles and the current modeled ones. We think
that this approach gives a very transparent and intuitive understanding of what the algorithm is doing. We note that the best results have been
found applying Eq. (\ref{eq:iteration_alpha}) every $\sim 3$ iterations of the LM algorithm.


\subsection{The synthesis penalty}
When $\mathbf{q}$ is fixed, Eq. (\ref{eq:problem_l0_synthesis}) becomes again a sparsity-constrained problem
for $\alphabold$. In this case, the sparsity penalty $h(\alphabold)=\Vert \alphabold \Vert_p$ is much simpler, but the function 
$g(\alphabold)=\chi^2_{\mathbf{W}^T}(\mathbf{q},\alphabold)$ is
more complex because of the presence of the synthesis operator $\mathbf{W}^T$. We propose to solve
the problem using the following first-order iterative scheme:
\begin{equation}
\alphabold_{i+1} = \mathrm{prox}_{p,\lambda,\tau} 
\left[\alphabold_i - \tau \nabla_{\alpha} \chi^2_\mathbf{W^T}(\mathbf{q}_{i+1},\alphabold_i) \right] 
\quad \text{with $\mathbf{q}_{i+1}$ fixed},
\label{eq:iteration_alpha2}
\end{equation}
where $\tau < 2 / \Vert \mathbf{H}_\alpha \Vert^2$, which can be computed from the spectral norm of the transformation
matrix $\mathbf{W}$, and the proximal operator $\mathrm{prox}_{p,\lambda,\tau}$ is described in the following. We point
out that a more complex second-order iterative scheme is discussed in the Appendix.

\subsection{Computing the proximal operators}
\label{sec:proximal}
The previous iterative schemes rely on the existence of algorithms for the computation 
of the proximal projection operators $\mathrm{prox}_{p,\lambda,\tau}(\alphabold)$ and $\mathrm{prox}_{p,\lambda,\mathbf{W}}(\alphabold)$.

\subsubsection{Computing $\mathrm{prox}_{p,\lambda,\tau}(\alphabold)$}
This operator is very simple to compute for 
typical choices of $p$ \citep{parikh_boyd14}. Useful cases of the regularization term are the $\ell_0$-norm ($p=0$) or 
the $\ell_1$-norm ($p=1$). In the case of the $\ell_0$-norm, the proximal
operator reduces to the hard-thresholding operator, which is trivially given by
\begin{equation}
\mathrm{prox}_{p=0,\lambda,\tau} (\alphabold) = 
\begin{cases}
\alphabold & \quad |\alphabold| > \tau \lambda \\
\mathbf{0} & \quad \mathrm{otherwise}.
\end{cases}
\label{eq:proximal_l0}
\end{equation}
where $\tau$ is the step-size defined in \S\ref{sec:prior} and $\lambda$ is the regularization parameter that we introduced in Eq.~\ref{eq:problem_l0}.
For the $\ell_1$-norm, it reduces to the soft-thresholding operator, which is given by
\begin{equation}
\mathrm{prox}_{p=1\lambda,\tau}(\alphabold) = \mathrm{sign}(\alphabold) (|\alphabold|-\tau \lambda)_+,
\label{eq:proximal_l1}
\end{equation}
where $(\cdot)_+$ denotes the positive part. Other proximal operators with analytical expressions
can be found in \cite{parikh_boyd14}.

\subsubsection{Computing $\mathrm{prox}_{p,\lambda,\mathbf{W}}(\alphabold)$}
The solution of Eq. (\ref{eq:iteration_alpha}) for any $\mathbf{W}$, either orthogonal or not, is slightly
more complicated and requires some elements of proximal calculus \citep{parikh_boyd14}. 
A general algorithm for the solution of the proximal projection of Eq. (\ref{eq:iteration_alpha})
has been developed by \cite{fadili09}, that we reproduce in Alg. \ref{alg:proximal} for completeness. We note that
the algorithm is just a simple iterative scheme that has been proven to converge to the 
solution provided the step size $\tau < 2 / \Vert \mathbf{W} \Vert^2$.

\begin{algorithm}[!t]
\KwData{$\mathbf{W}$ and $\alphabold$}
\KwResult{$\mathrm{prox}_{p,\lambda,\mathbf{W}}(\alphabold)$}
Initialization: $\mathbf{y}_1=\alphabold$ and suitable step $\tau < 2 / \Vert \mathbf{W} \Vert^2$\;
 \While{not converged}{
 1. $r=\alphabold - \mathbf{W} \mathbf{y}_{i}$ \;
 2. $\mathbf{s} = \tau \mathbf{y}_i + \mathbf{W}^T \mathbf{r}$ \;
 3. $\mathbf{y}_{i+1} = \alphabold - \tau^{-1} \mathbf{W} \left[ \mathbf{s} - \mathrm{prox}_{p,\lambda,\tau}(\mathbf{s}) \right]$ \;
 }
 \Return $\mathbf{y}_i$
 \caption{Algorithm for proximal projection, extracted from \cite{fadili09}.}
 \label{alg:proximal}
\end{algorithm}

It is interesting to point out that, when $\mathbf{W}$ is an orthogonal transform, Alg. \ref{alg:proximal} hugely simplifies and
the solution to Eq. (\ref{eq:iteration_alpha}) can be obtained with \citep{starck10}:
\begin{equation}
\alphabold = \mathbf{W}^T \left[ \mathrm{prox}_{p,\lambda,\tau} \left( \mathbf{W} \alphabold \right) \right].
\label{eq:proximal_wavelet}
\end{equation}
In other words, the vector $\alphabold$ is first transformed, it is then thresholded using the appropriate
proximal operator, and finally transformed back to the original domain. If the transformation is
unitary, so that $\mathbf{W}^T \mathbf{W}=\mathbf{1}$, the result
of the application of Eq. (\ref{eq:proximal_wavelet}) leaves the value of $\alphabold$ unchanged for $\lambda=0$.

Another interesting use of Eq. (\ref{eq:proximal_wavelet}) is that it is equivalent to the first iteration
of Alg. \ref{alg:proximal}. We carry out experiments in the following sections to verify if using this simple
approximation in the general case of non-orthogonal $\mathbf{W}$ gives good results.

\subsection{Selection of $\lambda$}
As we show in the next section, $\lambda$ has a strong impact on the sparsity of the final solution. When $\lambda$ is too small,
overfitting clearly appears. On the contrary, if $\lambda$ is large, the fit is typically of bad quality. Therefore,
the selection of $\lambda$ 
requires some fine-tuning, but it is possible to have an order of magnitude estimation using very simple arguments.
As noted in Eqs. (\ref{eq:proximal_l0}) and (\ref{eq:proximal_l1}), the thresholding happens for values of the
$\alphabold$ parameters larger than $\tau \lambda$. Therefore, if we want to avoid values of $\alphabold$ smaller than $\eta$,
then $\lambda \sim \eta \Vert \mathbf{W} \Vert^2$. For the orthogonal unitary transforms, we find that
the spectral norm is $\Vert \mathbf{W} \Vert=1$.

\begin{algorithm}[!t]
\KwData{Stokes profiles, model atmosphere, transform $\mathbf{W}$ and $k$.}
\KwResult{Regularized solution}
Initialization: $\mathbf{q}_0$ and $\alphabold_0=0$, first estimation of solution\;
 \While{not converged}{
 1. Compute gradient $\nabla_q \chi^2_\mathbf{1}(\mathbf{q}_i,\alphabold_i)$ and Hessian matrices $\mathbf{H}_q$\;
 2. Modify Hessians: $\hat{\mathbf{H}}_q=\mathbf{H}_q+ \beta \mathrm{diag}(\mathbf{H}_q)$\;
 3. Update $\mathbf{q}$ : $\mathbf{q}_{i+1} = \mathbf{q}_{i} - \hat{\mathbf{H}}_q^{-1} \nabla_q \chi^2_{\mathbf{1}}(\mathbf{q}_i,\alphabold_i)$ \;
 4. Update $\alphabold$: $\alphabold_{i+1} = \mathrm{prox}_{p,\lambda,\mathbf{W}} 
 \left[ \mathbf{S}_{\mathrm{atm}}(\mathbf{q}_{i+1})-\mathbf{O} \right]$ using Alg. \ref{alg:proximal} every $k$ iterations.
 }
 \Return $\mathbf{q}$, $\alphabold$
 \caption{Proximal Levenberg-Marquardt algorithm for analysis penalty.} 
 \label{alg:proximal_analysis}
\end{algorithm}

\subsection{Summary}
The full step-by-step algorithms are described in Algs. \ref{alg:proximal_analysis} and \ref{alg:proximal_synthesis}, 
together with Alg. \ref{alg:proximal} for the application of the proximal operator
described in step 4 of Alg. \ref{alg:proximal_analysis}. We want to clarify to any potential user of this method that 
the only real difference between current inversion codes for the Stokes parameters and our approach 
is in point 4 of Alg. \ref{alg:proximal_analysis}, together with the necessity to include the systematic effects in the
calculation of the gradient and Hessian with respect to the $\mathbf{q}$ variables. This step is very 
easy to carry out and can be implemented in any existing inversion code with only a few lines of programming.

\section{Examples}
\label{sec:examples}
In this section we demonstrate the capabilities of Algs. \ref{alg:proximal_analysis} and \ref{alg:proximal_synthesis} in the inversion of 
Stokes profiles using several transformation matrices $\mathbf{W}$ and two datasets.
For simplicity and for the purpose of clarity, we only focus on Stokes $I$, for which the 
systematic effects are usually more important. Throughout this section, we use the $\ell_0$-norm
as regularization, so we apply the proximal operator of Eq. (\ref{eq:proximal_l0}).
The codes used in this paper can be obtained from \texttt{https://github.com/aasensio/proxStokesSystematics}.

\begin{figure*}
\includegraphics[width=\textwidth]{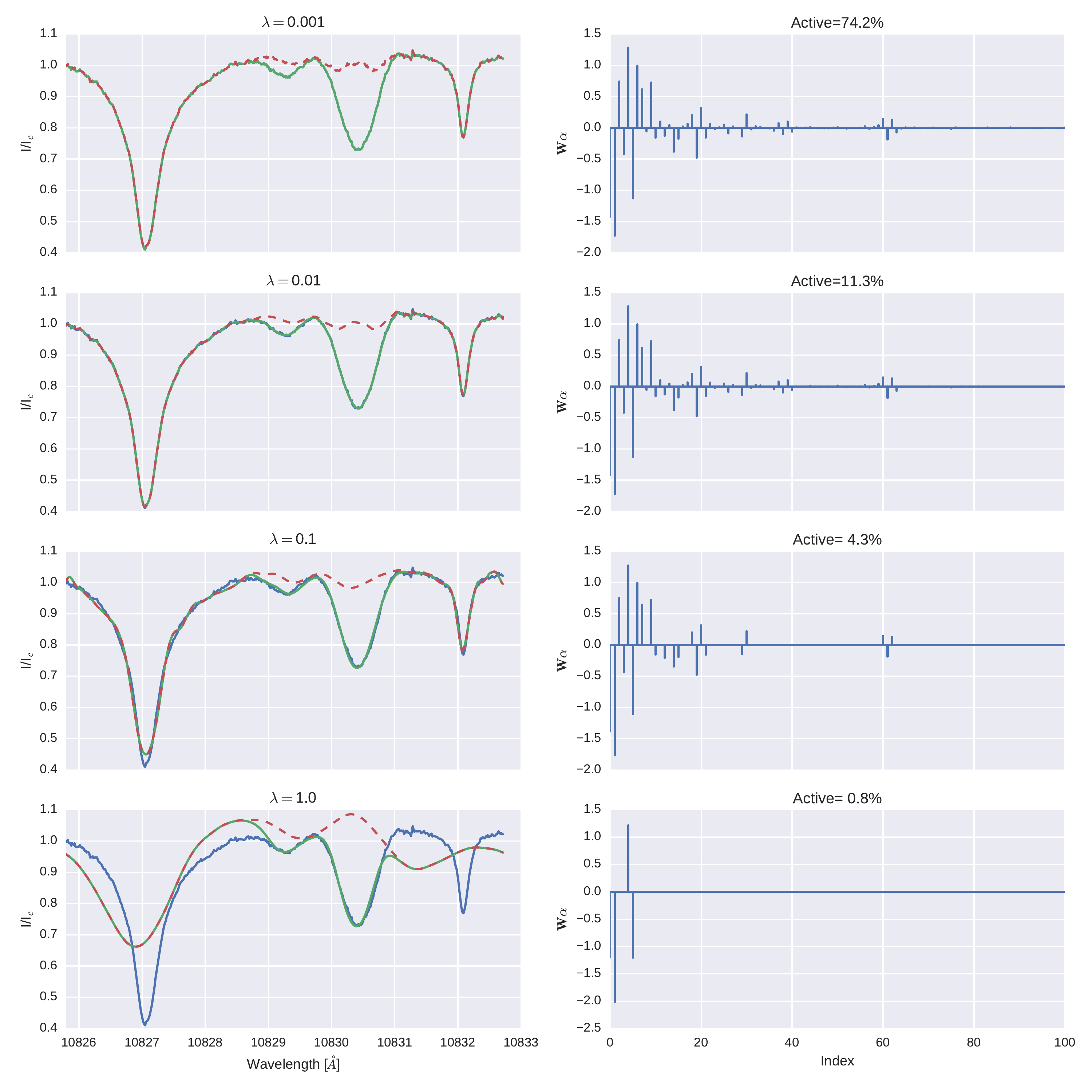}
\caption{Left panels: observed (blue) and fitted (green) Stokes $I$ profile for different values of the
regularization parameters $\lambda$ and using an orthogonal wavelet transform as the sparsity-inducing
transformation. The dashed red curve shows the inferred systematic effects. The spectral range corresponds to that around
the He \textsc{i} multiplet at 10830 \AA. Right panels: first 100 active wavelet coefficients of the set of
512 total coefficients. The percentage of active functions is shown in each panel.}
\label{fig:wavelet}
\end{figure*}

\subsection{Observations}
To show examples of application, we choose two spectral regions of special interest that suffer from
the problems described in this paper. The first one is the region around 10830 \AA, which 
contains the He \textsc{i} multiplet at 10830 \AA\ and is displayed in the 
right panel of Fig. \ref{fig:examples}. This multiplet is 
used for diagnosing the magnetic properties of chromospheric
material. Due to the potentially large Doppler shifts in the chromosphere \citep[e.g.,][]{lagg07}, it is interesting
to apply our approach to this multiplet.

The observations that we analyze have been obtained with the Vacuum Tower Telescope (VTT) on the 
Observatorio del Teide with the TIP-II instrument \citep{collados_tipII07}. The profiles have been 
extracted from a plage region close to a pore. The He \textsc{i} multiplet is 
synthesized with the help of \hazel\ \citep{asensio_trujillo_hazel08}, which gives the $\mathbf{S}_\mathrm{atm}(\mathbf{q})$ part of Eq. (\ref{eq:generative}). 
The spectral lines are characterized by the optical depth on the red component of the multiplet $\Delta \tau$,
the Doppler broadening of the line $v_\mathrm{th}$ and the bulk velocity of the plasma $v$.

The second region is the well-known region around 6301-6302 \AA\ that contains
an Fe \textsc{i} doublet. This region also contains two telluric absorptions. The Stokes profiles,
shown in the left panel of Fig. \ref{fig:examples}, have been obtained from the ground with the
POlarimetric LIttrow Spectrograph \citep[POLIS;][]{beck_polis05} and belong to the observations
of the quiet Sun analyzed by \cite{marian08}. 

To fit the profiles, we choose an atmosphere in local thermodynamic equilibrium and we infer the depth
stratification of the temperature, velocity along the line-of-sight and microturbulent
velocity. For this experiment we fix the number of nodes of the parameterization: 5 nodes for
the temperature, 3 for the bulk velocity and 1 for the microturbulent velocity. Like virtually
all inversion codes with depth stratification \citep[e.g.,][]{sir92}, we place the nodes equispaced in the $\log \tau_c$ axis,
with $\tau_c$ the continuum optical depth at 5000 \AA.

\subsection{Orthogonal wavelet regularization}
The orthogonal wavelet transform \citep{ripples01} is a very powerful 
sparsity-inducing transformation in cases in which the signal is smooth. One of the advantages
of the orthogonal wavelet transform is that a fast algorithm to compute the direct and inverse
transformation exists, without ever computing the transformation matrix $\mathbf{W}$. 

\begin{figure*}
\includegraphics[width=\textwidth]{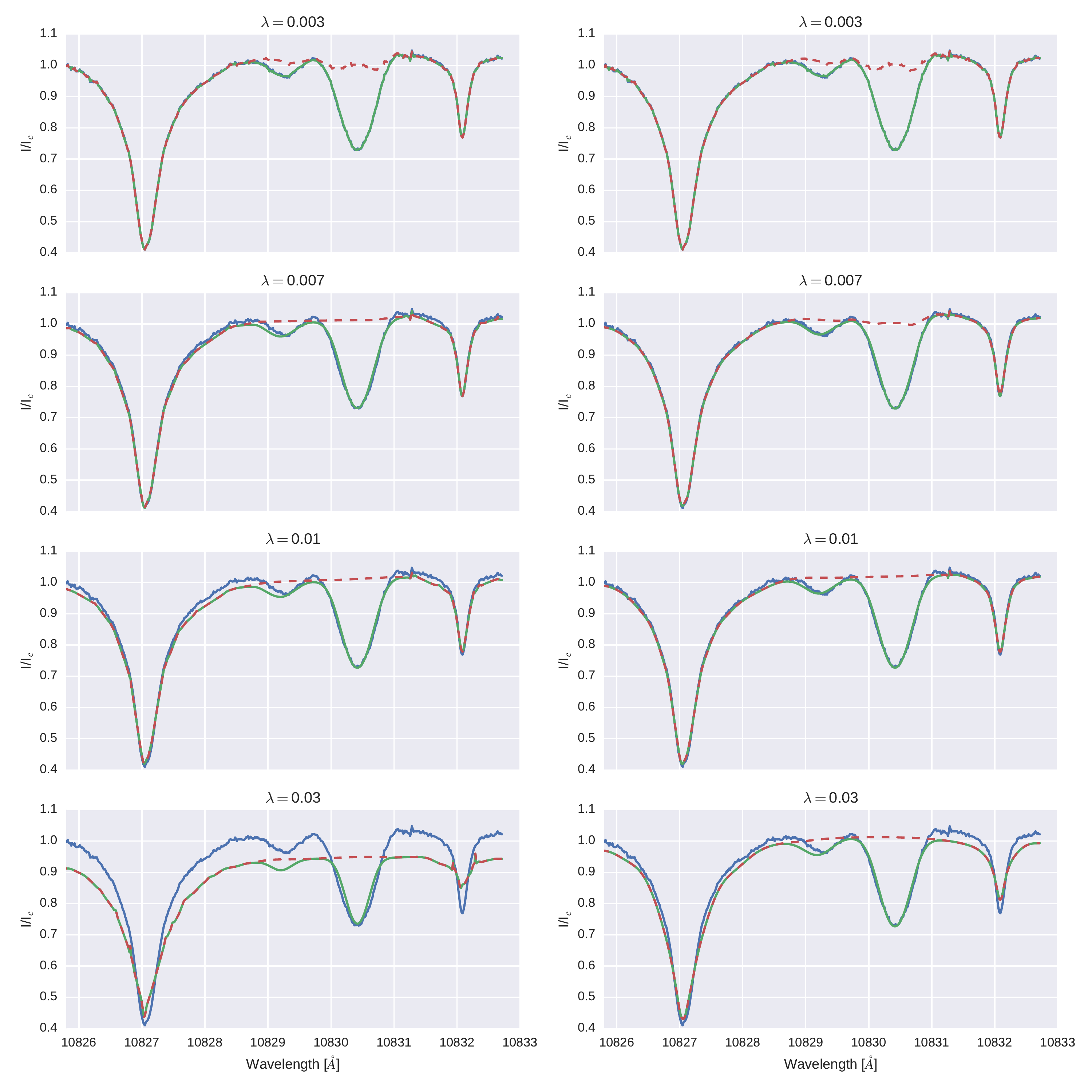}
\caption{Observed (blue), fitted (green) and systematic effects (dashed red) obtained using the IUWT
in the region around 10830 \AA\ and using the updated version of \hazel. The left panels
correspond to using the full Alg. \ref{alg:proximal}, while the right panel shows the results when using the
simplified Eq. (\ref{eq:proximal_wavelet}).}
\label{fig:iuwt_hazel}
\end{figure*}

\begin{algorithm}[!t]
\KwData{Stokes profiles, model atmosphere, transform $\mathbf{W}$ and $k$.}
\KwResult{Regularized solution}
Initialization: $\mathbf{q}_0$ and $\alphabold_0=0$, first estimation of solution\;
 \While{not converged}{
 1. Compute gradient $\nabla_q \chi^2_\mathbf{W^T}(\mathbf{q}_i,\alphabold_i)$ and Hessian matrices $\mathbf{H}_q$\;
 2. Modify Hessians: $\hat{\mathbf{H}}_q=\mathbf{H}_q+ \beta \mathrm{diag}(\mathbf{H}_q)$\;
 3. Update $\mathbf{q}$ : $\mathbf{q}_{i+1} = \mathbf{q}_{i} - \hat{\mathbf{H}}_q^{-1} \nabla_q \chi^2_\mathbf{W^T}(\mathbf{q}_i,\alphabold_i)$ \;
 4. Update $\alphabold$: $\alphabold_{i+1} = \mathrm{prox}_{p,\lambda,\tau}
 \left[ \alphabold_i - \tau \nabla_\alpha \chi^2_\mathbf{W^T}(\mathbf{q}_{i+1},\alphabold_i) \right]$\;
 }
 \Return $\mathbf{q}$, $\alphabold$
 \caption{Proximal Levenberg-Marquardt algorithm for synthesis penalty.} 
 \label{alg:proximal_synthesis}
\end{algorithm}

The specific approach for orthogonal transformations is irrelevant because both the
analysis and synthesis penalties are equivalent. For convenience, we choose to
do this study under the analysis penalty case.
We impose the sparsity constraint using the transformation matrix $\mathbf{W}$ associated with the 
Daubechies-8 orthogonal wavelet and for different values of the regularization parameter $\lambda$. 
The results are shown in Fig. \ref{fig:wavelet}. The left panel displays the observed data in 
blue, the final fit in green (including systematic effects) and the inferred systematic effects
in dashed red. The right panel shows the wavelet coefficients of the systematic effects, showing only the first
100 coefficients of the potential 512 coefficients (the wavelength axis contains 512 sampled points).
Each panel contains the percentage of active (non-zero) wavelet coefficients.
Note that the fit quality is strongly affected by the value of the regularization parameters $\lambda$.
The important point is that it is possible to find values of $\lambda$ that lead to a good fit of the 
He \textsc{i} multiplet and simultaneously fit the systematic effects. Of special relevance in this case is the 
extended wing of the Si \textsc{i}, which sometimes makes it difficult to set a continuum level 
for the 10830 \AA\ multiplet. Using this
approach, the continuum level is automatically obtained from the fit.

\begin{figure*}
\includegraphics[width=\textwidth]{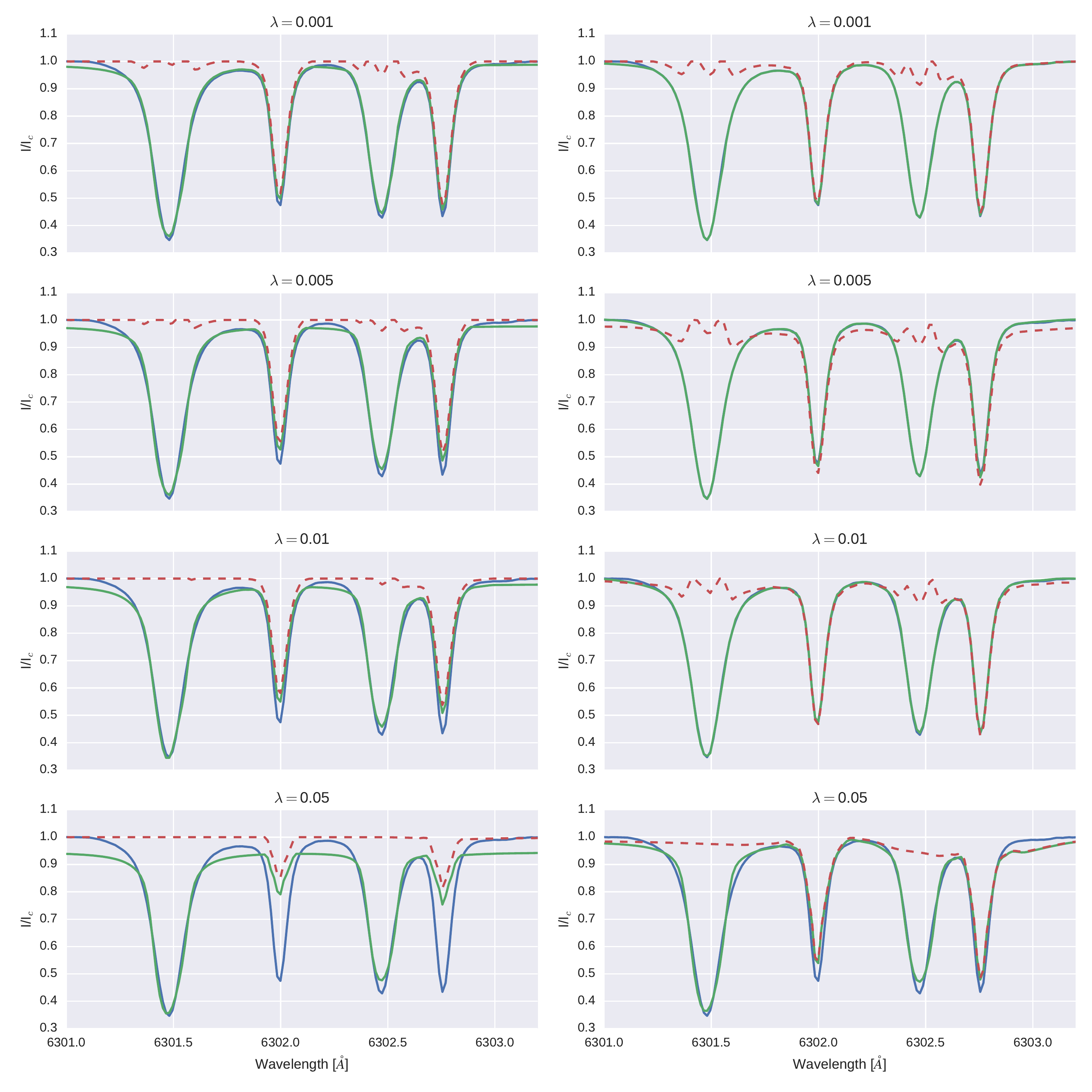}
\caption{Observed (blue), fitted (green) and systematic effects (dashed red) obtained using the IUWT
in the 6301-6302 \AA\ spectral region and using the inversion code that assumes local thermodynamic equilibrium. The left panels
correspond to using the full Alg. \ref{alg:proximal}, while the right panel shows the results when using the
simplified Eq. (\ref{eq:proximal_wavelet}).}
\label{fig:iuwt_lte}
\end{figure*}

When $\lambda$ is small, a strong overfitting of the data occurs, in which even the noise is
absorbed by the systematic effects. For very large $\lambda$, the method cannot fit the observations. 
For intermediate values of $\lambda$, a very nice fit is obtained. The optimal thresholding
of the wavelet coefficients is similar to the expected noise level, as pointed out by \cite{starck10}.
Using only 10\% of the wavelet coefficients is probably enough to have a fit of the whole spectral
region. It is interesting to note that the model that we impose for the He \textsc{i} multiplet is probably not enough for explaining
the observations at the noise level. This is the reason why some ``extra absorption'' is added by the systematic effects to the
wings of the red component. If this behavior is undesirable, it is potentially possible to introduce an extra
regularization to avoid them (for instance, an additional $\ell_0$ or $\ell_1$ regularization for the 
vector $\alphabold$). However, we demonstrate in the following that a better option is to regularize
using different $\mathbf{W}$ transformations.

\begin{figure*}
\includegraphics[width=\textwidth]{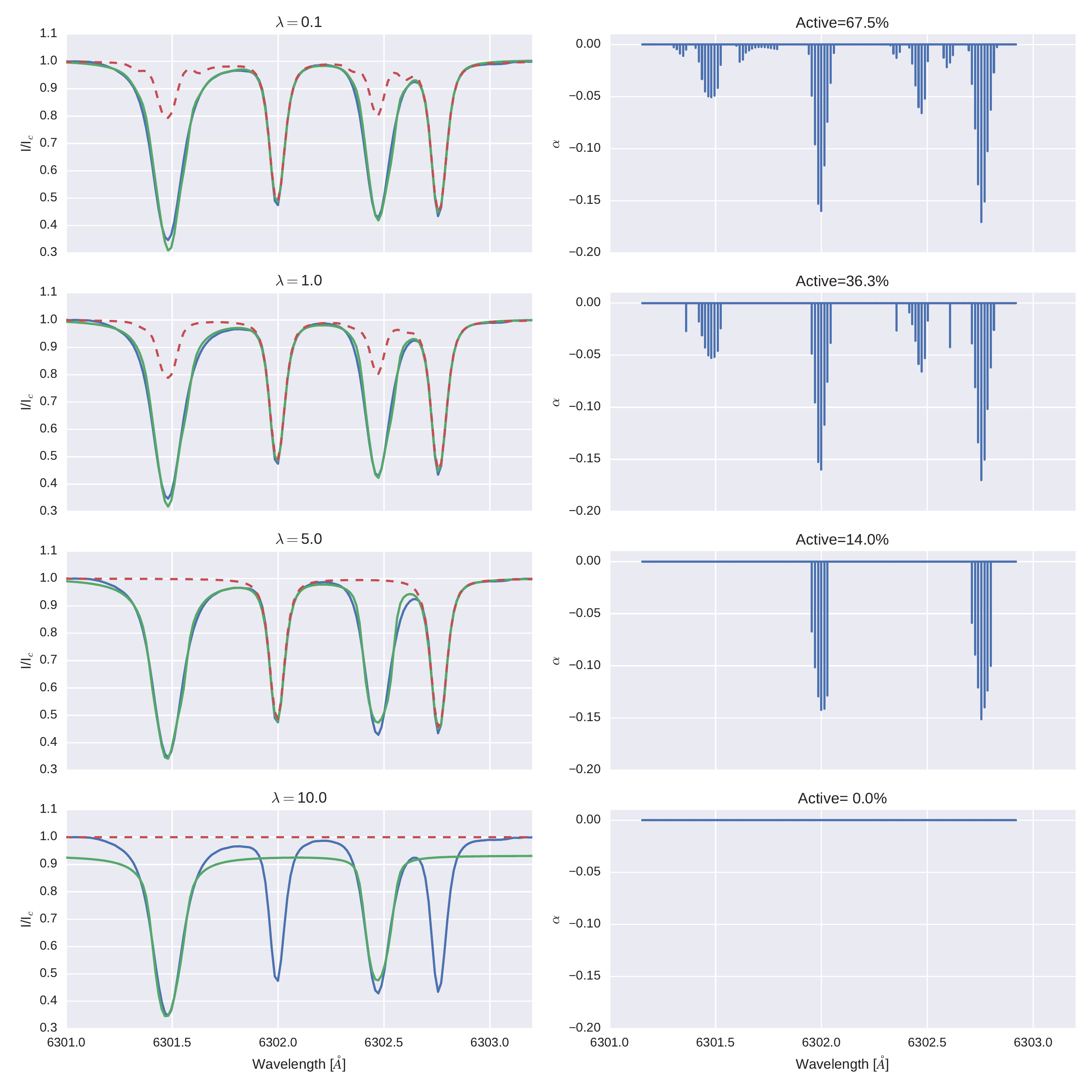}
\caption{Observed (blue) and fitted (green) Stokes $I$ profile for different values of the
regularization parameters $\lambda$ and using a dictionary made of Voigt functions centered
at every pixel. The dashed red curve shows the inferred systematic effects. The left panels show the 
results for the Fe \textsc{i} lines, while the right panels shows the active functions, with the
label indicating the percentage of active functions.}
\label{fig:lte_hazel_voigt}
\end{figure*}

\subsection{Isotropic undecimated wavelet regularization}
The isotropic undecimated wavelet transform \citep[IUWT;][]{starck94,starck10} algorithm is a non-orthogonal 
redundant multiscale transform that is well suited to objects that are more or less isotropic. This transform
has found great success for the denoising of astronomical images \citep[e.g.,][and references therein]{starck10}.
Recently, we have witnessed examples of applications to one-dimensional spectra \citep{machado13}.
Given that this is a redundant non-orthogonal transform, the analysis penalty approach is more
efficient from a computational point of view.

The IUWT can be efficiently applied using the \hbox{\emph{\`a-trous}} algorithm, which proceeds as follows. 
The original data $I(\lambda)$ are filtered, in our case using the B$_3$-spline given
by the filter $[1,4,6,4,1]/16$. The filtered data are substracted from the original ones, obtaining
what is commonly known as the \emph{detail}, $w_i(\lambda)$. The filtered data are again iteratively filtered (with scaled
filters) up to a depth $d$, computing the detail in each scale. The final smoothed signal will be
termed $c_d(\lambda)$. At the end, we have a smoothed version
of the original signal and a set of \emph{details} at each depth. The size of the IUWT is $d+1$ times
the original data, so the information is really encoded on the correlation among the transformed coefficients.
The data are reconstructed from the transformed data simply using:
\begin{equation}
I(\lambda) = c_d(\lambda) + \sum_{i=1}^d w_i(\lambda).
\end{equation}

The results shown in Fig. \ref{fig:iuwt_hazel} for the He \textsc{i} multiplet and \hbox{Fig. \ref{fig:iuwt_lte}}
for the Fe \textsc{i} doublet have been obtained using the IUWT up to $d=6$ and
applying the hard thresholding operator to all the details. The left panel shows the results when
the full \hbox{Alg. \ref{alg:proximal}} is used to apply the proximal algorithm, while the right
panels give the result when only the first iteration of this algorithm is used, which corresponds
to using Eq. (\ref{eq:proximal_wavelet}). The application of Eq. (\ref{eq:proximal_wavelet}) cannot
assure convergence to the correct solution to the problem, but we have tested that it
does a very good job. This simplifies the application of our proposed algorithm to any
existing inversion code because Eq. (\ref{eq:proximal_wavelet}) can be implemented in
only one or a few lines of code. 

When the value of $\lambda$ is small, we witness again the overfitting, which also includes
the addition of some broadening of the red component of the He \textsc{i}. When $\lambda$ is
large, the fit is of low quality. In the intermediate range close to the expected noise
standard deviation, we find an excellent fit of the whole profile, with a flat contribution
exactly where the He \textsc{i} multiplet is located.

It is interesting to note that we have empirically found that applying Eq. (\ref{eq:proximal_wavelet}) instead of solving
the full problem via Alg. \ref{alg:proximal} gives better and more robust results (see Figs. \ref{fig:iuwt_hazel} and \ref{fig:iuwt_lte}).
The reason has to be found on the fact that we are solving the optimization problem of Eqs. (\ref{eq:problem_l0}) or
(\ref{eq:problem_l0_synthesis}) by separating
it into two simpler problems via an alternating optimization method. It is a general characteristic
of these methods that they work better in practice if none of the two problems is solved with 
full precision at each iteration, but only approximately. If any of the two problems is solved with precision in
any iteration, one can produce some amount of overfitting that is very difficult to compensate
for in later iterations.

\subsection{Voigt functions}
The final example uses another non-orthogonal and redundant transformation, in this case not of general
applicability, but tailored to explain the systematic effects. The Hermite functions described
by \cite{hermite_deltoro03} is a good option, but we prefer to utilize a basis set made of Voigt functions centered
at every sampled wavelength point:
\begin{equation}
\phi_i(\lambda) = H \left( \frac{\lambda-\lambda_i}{\Delta}, a\right).
\end{equation}
Even though the basis is non-orthogonal, it is much easier in this case to work in
the synthesis prior approach, because the results are much more transparent. The reason
is that we directly impose the sparsity constraint on the coefficients associated with
all the Voigt functions.

In this case, we focus on the 6301-6302 \AA\ spectral region. The systematic effects are modeled with Voigt functions 
placed at every sampled wavelength between 6301.15 and 6302.93 \AA, which includes both Fe \textsc{i} lines and
both telluric contaminations. The damping constant is fixed to 
$a=0.8$, which gives line wings similar to those observed in the telluric lines. To boost sparsity, 
we consider two different widths, $\Delta=14.8, 29.6$ m\AA, although more fits and damping parameters 
can be easily considered with ease. 

The results are displayed in Fig. \ref{fig:lte_hazel_voigt} for different values of $\lambda$. The
left panel shows the observations and the fit, together with the inferred systematic effects. The right panel shows the active
Voigt functions. When $\lambda$ is too small, a very good fitting is found, but we can see that the optimization induces 
corrections on the spectral lines of interest. This is, in principle, undesired, although it is
clearly stating that the model proposed for the spectral lines cannot fit them with enough
precision. When $\lambda$ is large, the fit of the systematic effects is bad, which negatively impacts 
the fit of the spectral lines of interest. For intermediate values of the regularization parameter, we find 
an acceptable fit, where only the telluric absorption lines are fitted by the systematic effects with a very
sparse solution.



\section{Conclusions}
In the present study we have developed a new method that allows to deal with systematic effects in data inversions. 
These systematic effects can include a variety of calibration defects, spectral lines from the Earth's atmosphere or spectral features that
are present in the observations that our model atmosphere cannot reproduce. Our method builds upon the assumption that these
systematic effects are not random noise and therefore they have an inherent \emph{level of sparsity} in
a suitable basis set.

We have shown that the assumption of sparsity induces a very powerful regularization because the resulting cross-talk between the 
inversion model and the systematic effects can be made very low. To do so, we have introduced a regularization parameter $\lambda$ that
must be adjusted according to the needs of each problem based on experience. However, we argue that this step is not very different 
from selecting the weights of the observations in current implementations of inversions codes and a rough estimate can be obtained \emph{a-priori}.

Although the mathematical characterization of the algorithm can appear a bit convoluted and depends on
recent advances on the optimization of non-smooth functions, a few extra lines of code should suffice to make
it work with current inversion codes. Our experience with the SIR and Hazel codes has positively confirmed it. When these systematic effects are not 
corrected, inversion codes usually try to overcompensate their effect by converging the parameters of the model to a wrong solution. Our method 
minimizes the impact of such effects in the convergence of the solution when an adequate value of the regularization parameter is selected.

Our method has a few interesting advantages over the downweighting technique 
for dealing with systematic effects. Arguably the most important one concerns the assumptions
imposed on the generative model. Using the $\chi^2$ merit
function requires that the uncertainty of the residual between the observations
and the model is Gaussian with zero mean and a certain variance. This is not true unless one is
able to \emph{model all} expected signals. Only if everything is modeled (to a certain level, of course), one 
is sure that the results can be interpreted appropriately (for instance, error bars). A second advantage
concerns the reduction in the subjectivity on the election of parameters. One only needs to choose
the value of $\lambda$ and the method automatically adapts the solution to the systematic effects.

%

\appendix
\section{Second-order iterative scheme in the synthesis prior}
It is possible to use a second-order iterative scheme that would improve over Eq. (\ref{eq:iteration_alpha2})
when using the synthesis prior approach. Instead of a simple step $\tau$, one uses the full Hessian, so that
the iterative scheme is updated to read \citep{lee12}:
\begin{align}
\alphabold_{i+1} &= \mathrm{prox}_{p,\lambda}^{H_\alpha} \left[ \alphabold_{i} - \mathbf{H}_\alpha^{-1} \nabla_\alpha
\chi^2_\mathbf{W^T}(\mathbf{q}_{i+1},\alphabold_i) \right] \quad \text{with $\mathbf{q}_{i+1}$ fixed}.
\end{align}
The main obstacle is the computation of the so-called scaled proximal operator $\mathrm{prox}_{p,\lambda}^{H_\alpha}(\alphabold)$.
It can be demonstrated that this proximal operator can be efficiently computed using the accelerated
FISTA iteration \citep{beck_teboulle09} that is displayed in Alg. \ref{alg:scaled_proximal}.

\begin{algorithm}[!t]
\KwData{$\mathbf{H}_\alpha$ and $\alphabold$}
\KwResult{$\mathrm{prox}_h^H(\alphabold)$}
Initialization: $t_0=1$, $\mathbf{y}_1=\alphabold$ and step $\tau < 2 / \Vert \mathbf{H}_\alpha \Vert^2$\;
 \While{not converged}{
 1. Update values: $\mathbf{y}'_{i+1} = \mathbf{y}_i - \tau \mathbf{H}_\alpha (\mathbf{y}_i-\alphabold)$ \;
 2. Apply projection: $\alphabold_{i+1}=\mathrm{prox}_{p,\lambda,\tau}(\mathbf{y}'_{i+1})$\;
 3. Update weight: $t_{i+1}= \frac{1+\sqrt{1+4t_i^2}}{2}$\;
 4. Update solution: $\mathbf{y}_{i+1} = \mathbf{y}_{i} + \left( \frac{t_i-1}{t_{i+1}} \right) (\mathbf{y}'_{i+1} - \mathbf{y}_{i})$\;
 }
 \Return $\mathbf{y}_i$
 \caption{FISTA algorithm for scaled proximal projection.} 
 \label{alg:scaled_proximal}
\end{algorithm}

\begin{acknowledgements}
Financial support by the Spanish Ministry of Economy and Competitiveness 
through projects AYA2014-60476-P, AYA2014-60833-P and Consolider-Ingenio 2010 CSD2009-00038 
are gratefully acknowledged. JdlCR is supported by grants from the
Swedish Research Council (Vetenskapsr\aa det) and the Swedish National Space Board.
AAR also acknowledges financial support through the Ram\'on y Cajal fellowship. 
This research has made use of NASA's Astrophysics Data System Bibliographic Services.
This article is based on observations taken with the VTT telescope operated on the island of 
Tenerife by the Kiepenheuer-Institut für Sonnenphysik in the Spanish Observatorio del Teide of the Instituto de Astrofísica de Canarias (IAC).
We acknowledge the community effort devoted to the development of the following open-source Python packages that were
used in this work: \texttt{numpy}, \texttt{matplotlib}, \texttt{scipy}, \texttt{seaborn} and \texttt{PyWavelets}.
\end{acknowledgements}


\end{document}